\documentclass{rnaastex}

\begin{document}

\title{Ejection of material --"Jurads" -- from post main sequence planetary systems}

\correspondingauthor{Brad Hansen}
\email{hansen@astro.ucla.edu, ben@astro.ucla.edu}

\author{Brad Hansen}
\affiliation{Department of Physics and Astronomy, UCLA}

\author{Ben Zuckerman}
\affiliation{Department of Physics and Astronomy, UCLA}


\section{} 

The identification of (1I/2017 U1) 'Oumuamua as the first known minor body of
extrasolar origin (Williams 2017; de la Fuente Marcos \& de la Fuente Marcos 2017) 
has generated substantial interest  
(e.g. Cuk 2017; Laughlin \& Batygin 2017; Raymond et al. 2017; Jackson et al. 2017).
One expects that
the assembly of planetary systems is an inherently dynamic process, in which
substantial amounts of material are ejected into interstellar space. As such,
 'Oumuamua represents the harbinger of a bounty to come, as more  
such objects are discovered by surveys such as Pan-STARRS and LSST. This 
will allow us to probe the small body constituents of other planetary systems.

However this population is not the first probe of extrasolar minor bodies,
 as observations of the heavy element pollution of white dwarfs (Zuckerman et al. 2003, 2007;
Jura \& Young 2014; Farihi 2016 and references therein) has already provided
a  wealth of exogeological information. In this note we point out that
these two populations may well be linked.

The consensus model for pollution of white dwarf atmospheres is that, typically, the 
heavy elements stem from asteroidal bodies that have been tidally disrupted upon close 
passage to a white dwarf. To get close enough, minor bodies must be perturbed onto 
almost radial orbits via scattering by planets at several AU (e.g. Debes \& Sigurdsson 
2002; Debes, Walsh \& Stark 2012; Frewen \& Hansen 2014; Mustill et al. 2017). Such a 
scattering process will, inevitably, eject material into interstellar space.

To estimate this mass, we turn to the white dwarf sample observed by Jura \& Xu 
(2012).
Summing the accretion rates of all observed stars and averaging over the entire,
approximately complete, sample, 
we infer an average rate of accretion of rocky material, per white dwarf,
$\sim 3 \times 10^8$ g/s.
Adopting a median age of $\sim 5$~Gyr for white dwarfs in the solar neighbourhood,
this implies that approximately $0.01 M_{\oplus}$ of material is accreted per
local white dwarf. The relative fraction of 
ejected to accreted mass depends on the mass and eccentricity of
the scattering planet (Frewen \& Hansen 2014) but values of $\sim $10--100 
are expected for planets in the Saturn--Jupiter range and moderate orbital 
eccentricities ($e<0.3$). 
This implies that 0.1--1.0$M_{\oplus}$ of material is ejected
per white dwarf. To convert this into a global interstellar density,
we note that white dwarfs make up $\sim 0.06$ of the stars in the local 10 parsec
sample (Henry et al. 2006) so that the ejected material amounts to as much as $\sim 0.06 M_{\oplus}$ per 
star.
Although this is only a fraction of the mass per star inferred from the properties of 
Oumuamua (e.g. Cuk 2017), the uncertainties in such estimates suggest any process that 
produces numbers within an order of magnitude is worthy of consideration.

Another reason to consider evolved planetary systems as a significant source of 
interstellar minor bodies is lack of evidence of a coma emanating from Oumuamua 
(Jewitt et al. 2017). Post main sequence evolution of Sun-like stars  
substantially increases the illumination experienced by distant small bodies in 
surrounding planetary systems. Models for G, F and K stars 
(Paxton et al. 2013) suggest that all bodies with starting orbits within 100--200~AU 
will be heated above equilibrium temperatures of 150~K during the giant phase of 
stellar evolution. Thus, most of the ejected material described above should be 
depleted in volatiles, consistent with the properties of material 
accreted by white dwarfs (Jura \& Young 2014; Farihi 2016).  Although the evidence 
that Oumuamua is asteroidal rather than cometary is tentative (Jewitt et al. 2017; Fitzsimmons et al. 2017), 
this does indicate that volatile-depleted material need not result 
from deep in the potential wells of main sequence stars.

The population of interstellar comets and asteroids promises to reveal much 
about planetary systems that orbit on large scales around nearby stars, but care is
needed to delineate the  multiple potential pathways that could
contribute to the overall population. The existence of polluted white dwarfs is direct 
evidence that post main sequence systems remain dynamically active, and our estimates 
suggest that they can make a non-negligible contribution to the observations, 
especially of volatile-depleted bodies.

Finally, we propose the name ''Jurads'' to denote bodies drawn from this evolutionary path, in 
honor of our late colleague Mike Jura, who pioneered studies of this kind.  




\begin{thebibliography}{99}

\bibitem[Cuk(2017)]{C17} Cuk, M., 2017, arXiv:1712.01823

\bibitem[Debes.Sigurdsson(2002)]{DS02} Debes, J. H. \& Sigurdsson, S., 2002, \apj, 572, 556

\bibitem[Debes(2012)]{DWS12} Debes, J. H., Walsh, K. J. \& Stark, C., 2012, 747, 148

\bibitem[delfm2(2017)]{dlF2017} de la Fuente Marcos, C. \& de la Fuente Marcos, R., 2017, arXiv:1711.00445

\bibitem[Farihi(2016)]{F16} Farihi, J., 2016, New Astron. Reviews, 71, 9

\bibitem[Fitz(2017)]{Fitz17} Fitzsimmons, A.,  et al., 2017, arXiv:1712.06552

\bibitem[Frewen.Hansen(2014)]{FH14} Frewen, S. \& Hansen, B., 2014, MNRAS, 439, 2442

\bibitem[Henery(2006)]{H06} Henry, T. J, Jao, W.-C., Subasavage, J. P.,  et al., 2006, \aj, 132, 2360

\bibitem[Jacosn(2017)]{JTH17} Jackson, A. P., Tamayo, D., Hammond, N., et al., 2017, arXiv:1712.04435

\bibitem[Jewitt(2017)]{JLR17} Jewitt, D., Luu, J., Rajagopal, J. et al., 2017, \apj, 850, L36

\bibitem[Jura.Young(2014)]{JY14} Jura, M. \& Young, E. D., 2014, AREPS, 42, 45

\bibitem[Jura.Xu(2012)]{JX12} Jura, M. \& Xu, S., 2012, \aj, 143, 6

\bibitem[LB(2017)]{LB17} Laughlin, G., \& Batygin, K., 2017, arXiv:1711.02260

\bibitem[MVV(2017)]{MVV17} Mustill, A. J., Villaver, E., Veras, D., et al., 2017, arxiv:1711.02940

\bibitem[Pax(2013)]{P13} Paxton, W. B., Cantiello, M., Arras, P., et al., 2013, \apjs, 208, 4

\bibitem[Ray(2017)]{R17} Raymond, S. N., Armitage, P. J., Veras, D., et al., 2017, arXiv:1711.09599

\bibitem[Williams(2017)]{W17} Williams, G.\ 2017 Minor Planet Center Electronic Circular 2017-U181 

\bibitem[Zuckerman(2003)]{Z03} Zuckerman, B., Koester, D., Reid, I.N. et al.\ 2003, \apj, 596, 477

\bibitem[Zuckerman(2007)]{Z07} Zuckerman, B., Koester, D., Melis, C. et al., 2007, \apj, 671, 872

\end{thebibliography}
\end{document}